\documentclass[aps,prb,twocolumn,amsmath,amssymb,superscriptaddress,floatfix]{revtex4}
\usepackage{graphicx}
\usepackage{bm}
\usepackage[usenames]{color}
\bibstyle{apsrev.bib}

\newcommand{\be}{\begin{equation}}
\newcommand{\ee}{\end{equation}}
\newcommand{\beqn}{\begin{eqnarray}}
\newcommand{\eeqn}{\end{eqnarray}}

\begin{document}

\title{Corner contribution to percolation cluster numbers}

\author{Istv\'an A. Kov\'acs}
\email{kovacs.istvan@wigner.mta.hu}
\affiliation{Wigner Research Centre, Institute for Solid State Physics and Optics,
H-1525 Budapest, P.O.Box 49, Hungary}
\author{Ferenc Igl\'oi}
\email{igloi.ferenc@wigner.mta.hu}
\affiliation{Wigner Research Centre, Institute for Solid State Physics and Optics,
H-1525 Budapest, P.O.Box 49, Hungary}
\affiliation{Institute of Theoretical Physics,
Szeged University, H-6720 Szeged, Hungary}
\author{John Cardy}
\email{j.cardy1@physics.ox.ac.uk}
\affiliation{Rudolph Peierls Centre for Theoretical Physics, University of Oxford, 1 Keble
Road, Oxford OX1 3NP, United Kingdom}
\affiliation{All Souls College, Oxford}
\date{\today}


\begin{abstract}
We study the number of clusters in two-dimensional ($2d$) critical percolation, $N_{\Gamma}$, which
intersect a given subset of bonds, $\Gamma$. In the simplest case, when $\Gamma$ is a simple closed
curve, $N_{\Gamma}$ is related to the
entanglement entropy of the critical diluted quantum Ising model, in which $\Gamma$
represents the boundary between the subsystem and the environment. Due to corners in $\Gamma$
there are universal logarithmic corrections to $N_{\Gamma}$, which are calculated in
the continuum limit through conformal invariance, making use of the Cardy-Peschel formula.
The exact formulas are confirmed by
large scale Monte Carlo simulations. These results are extended to anisotropic percolation where they confirm a result of discrete holomorphicity.
\end{abstract}

\maketitle
\section{Introduction}
\label{sec:intr}
In the percolation process sites or bonds of a regular lattice (or a graph) are independently open
with a probability, $p$ and one is interested in the statistics of clusters of connected
sites. In the thermodynamic limit for dimensions $d \ge 2$ there is a critical point at $p=p_c$,
at which a second order phase transition takes place. Critical percolation is scale invariant and its scaling limit is also believed to be conformally invariant. This has been rigorously proved for some lattices\cite{smirnov}. Conformal
invariance has been used to predict different relations about the order-parameter profiles,
correlation functions, crossing probabilities, etc. of critical percolation. Some of the
conformal results have been subsequently derived by rigorous mathematical methods, such as
by  Schramm-Loewner evolution\cite{sle}.

In percolation the questions one usually asks concern the distribution of clusters, or the
properties of the largest ones, but comparatively less attention is paid to the total number
of clusters, which is expected to scale with the volume of the system under consideration.
To the leading volume term there are corrections due to boundary effects, such as surfaces,
edges and corners. The total number of clusters and its correction terms are generally
non-universal, thus they depend on the type of percolation (bond or site) and the details
of the  lattice. The only exception could be the corner contribution, which we are
going to study in this paper.

To be specific we consider bond percolation in a $2d$ square lattice and denote by $\Gamma$ a
subset of bonds. Initially we take this to be a simple closed curve consisting of straight edges
and corners and will generalize later. We are interested in the number of clusters, $N_{\Gamma}$,
which intersect $\Gamma$, in the scaling limit when $\Gamma$ is large but still much smaller than the total size of the system. This problem is closely related to the
entanglement entropy of the bond-diluted quantum Ising model which is defined
in the same lattice and $\Gamma$ represents the interface, which separates a subsystem from the
rest of the system.

In the following we calculate $N_{\Gamma}$ by different methods, both analytically and numerically.
In the scaling limit we consider
the random cluster representation of the $Q$-state Potts model and in the $Q \to 1$ limit
we calculate $N_{\Gamma}$. In particular we obtain a
logarithmically divergent contribution due to corners and its prefactor is calculated through conformal invariance.
Using different geometries of $\Gamma$ we have performed large scale
Monte Carlo simulations, the results of which are confronted with the conformal prediction.
We have also considered anisotropic percolation, in which the probabilities $(p_x,p_y)$ that horizontal and vertical bonds are open are different.

\section{Potts model representation}
\label{sec:Potts}
Bond percolation is well known to be related to the $Q \to 1$ limit of the $Q$-state Potts model, 
defined on a lattice with sites $i=1,2,\dots,n$ and $m$ nearest neighbor bonds.
The partition function is given by
\be
Z(Q)=\sum_{s} \prod_{\left\langle ij\right\rangle } \exp \left( K \delta_{s_i,s_j} \right)\;,
\label{Potts}
\ee
where the sum is over all spin configurations, $s=(s_1,s_2,\dots,s_n)$, $s_i=1,2,\dots, Q$,
the product runs over all nearest neighbor bonds, $\delta_{s_i,s_j}$
is the Kronecker symbol and $K$ is the reduced coupling, which is the ratio of the pair interaction and the temperature.
Using the identity: $\exp\left( K \delta_{s_i,s_j} \right)=1+\frac{p}{1-p} \delta_{s_i,s_j}$
with $p=1-e^{-K}$, the sum of products in $Z$ is written in terms of the so called Fortuin-Kasteleyn
clusters\cite{Fortuin-Kasteleyn}, denoted by $F$. In $F$ the edge of the lattice  $i,j$ is occupied, if
a factor $\frac{p}{1-p} \delta_{s_i,s_j}$ is present and
in any connected cluster the spins are in the same orientation. 
For a given element of $F$ there are $N_{tot}(F) \le n$ connected components and $M(F) \le m$
occupied bonds so that the partition function reads:
\beqn
&Z(Q)=\sum_{F} Q^{N_{tot}(F)}\left( \dfrac{p}{1-p}\right)^{M(F)} \cr
&\sim \sum_{F} Q^{N_{tot}(F)} p^{M(F)} {(1-p)}^{m-M(F)}
=\left\langle Q^{N_{tot}}\right\rangle\;.
\label{Z}
\eeqn
In this random cluster representation $Q$ is a real parameter and percolation is recovered
in the $Q \to 1$ limit, when the critical point in the square lattice is given by $p_c=1/2$.
The mean total number of clusters in percolation is:
\be
\left\langle N_{tot}\right\rangle=\left. \dfrac{\partial \ln Z(Q)}{\partial Q}\right|_{Q=1}\;.
\label{N_tot}
\ee
Now impose a boundary condition that all the Potts spins are fixed (say, in the state $1$) on $\Gamma$.
The partition function is now:
\be
Z_{\Gamma}(Q) \sim \left\langle Q^{N_{tot}-N_{\Gamma}}\right\rangle\;,
\label{Z_Gamma}
\ee
where $N_{\Gamma}$ is the number of clusters which intersect $\Gamma$. Hence
\be
\left\langle N_{tot} - N_{\Gamma} \right\rangle=\left. \dfrac{\partial \ln Z_{\Gamma}(Q)}{\partial Q}\right|_{Q=1}\;.
\label{N_tot-N_Gamma}
\ee
On the other hand at the critical point $p=p_c$ we can write:
\beqn
\ln Z(Q)&\sim&A f_b(Q) \cr
\ln Z_{\Gamma}(Q)&\sim&A f_b(Q)+L_{\Gamma} f_s(Q) + C_{\Gamma}(Q) \ln L_{\Gamma}\;,
\label{log_Z}
\eeqn
where $A \propto n$ is the total area, $L_{\Gamma}$ is the length of $\Gamma$, and $f_b$ and
$f_s$ are the bulk and surface free-energy densities, respectively, which are non-universal.
The last term represents the corner contribution.
Hence we obtain:
\be
\left\langle N_{\Gamma} \right\rangle=-f'_s(1)L_{\Gamma}-C'_{\Gamma}(1)\ln L_{\Gamma} \;.
\label{N_Gamma}
\ee
Cardy and Peschel\cite{cardypeschel} considered these corner contributions to the free energy of a general conformally invariant system in domains with a boundary. Their results apply equally to an exterior boundary, with corners with interior angle $\gamma_k$, or an interior boundary, with $\gamma_k$ replaced by $2\pi-\gamma_k$. An important property of percolation is its locality, that is the partition function $Z_\Gamma(Q)$ is exactly the product of the partition function $Z^{\rm int}_\Gamma(Q)$ for the interior of $\Gamma$, and $Z^{\rm ext}_\Gamma(Q)$ for the exterior. Thus we may apply the Cardy-Peschel result to each of these. Note that this would not be correct if we had given the clusters intersecting $\Gamma$ a weight $\not=1$. 
We therefore find from Ref.~\onlinecite{cardypeschel} that the prefactor of the logarithm in Eq.(\ref{log_Z}) is
given by:
\beqn
C_{\Gamma}(Q)&=&\dfrac{c(Q)}{24} \sum_k \left[ \left( \dfrac{\pi}{\gamma_k}\right)- \left( \dfrac{\gamma_k}{\pi}\right)\right. \cr
&+&\left. \left(\dfrac{\pi}{2 \pi-\gamma_k}\right)-\left(\dfrac{2 \pi-\gamma_k}{\pi}\right)\right]\;,
\label{cardy_peschel}
\eeqn
where $\gamma_k$ is the interior angle at each corner, and $c(Q)$ is the central charge of the $Q$-state Potts model, and the two sets of terms come from the interior and exterior contribution\cite{note}.

Using\cite{CG} $c=(6-\kappa)(3 \kappa -8)/2 \kappa$ and $\sqrt{Q}=-2 \cos(4 \pi/\kappa)$ we have:
\be
c'(1)=\dfrac{5 \sqrt{3}}{4 \pi}\;
\label{c'}
\ee
The corner contribution to $\left\langle N_{\Gamma} \right\rangle$ which is derived here for bond
percolation is expected to be universal, thus to be valid for site percolation, too.
We are going to compare the conformal results in Eqs.(\ref{N_Gamma},\ref{cardy_peschel})
and (\ref{c'}) with those of numerical calculations for different forms of $\Gamma$ in Sec.\ref{sec:numerical}.

\section{Entanglement entropy of the diluted quantum Ising model}
\label{sec:entropy}

The problem studied in Sec.\ref{sec:Potts} for percolation is closely related
to the entanglement entropy of a bond-diluted quantum Ising model and here we follow Refs.~\onlinecite{lin07,yu07} .
The model is defined by the Hamiltonian:
\be
{\cal H}=-\sum_{\left\langle ij\right\rangle } J_{ij} \sigma_i^x \sigma_j^x -\sum_i h \sigma_i^z\;,
\label{Ising}
\ee
in terms of the $\sigma_i^{x,z}$ Pauli matrices at site $i$. The first sum in Eq.(\ref{Ising})
runs over nearest neighbors and the $J_{ij}$ coupling equals $J>0$ with probability $p$ and
equals $J=0$ with probability $1-p$. At the percolation transition point, $p_c$, for small
transverse field, $h$, there is a line of phase transition the critical properties of
which are controlled by the percolation fixed point\cite{senthil_sachdev}. The ground state of ${\cal H}$
is given by a set of ordered clusters, which are in the same form as for percolation. Now consider
a subsystem, $A$ the boundary of which is represented by $\Gamma$ and calculate the entanglement
entropy between the subsystem and the environment, which is given by
${\cal S}_{\Gamma}=-{\rm Tr}(\rho_A \log_2 \rho_A)$ in terms of the reduced density matrix $\rho_A$.
Here we note, that in each ordered cluster the spins are in a maximally entangled GHZ
state, thus
for a given realization of disorder in the small $h$ limit all those clusters give a unit (1) contribution
to the entanglement entropy, which intersect $\Gamma$ and contain also at least one point of the environment.
We shall call them crossing clusters, thus ${\cal S}_{\Gamma}$ is given by the number of crossing
clusters and we are interested in their average value: $\left\langle {\cal S}_{\Gamma}\right\rangle$.
Note that ${\cal S}_{\Gamma} \le N_{\Gamma}$, and thus $\left\langle {\cal S}_{\Gamma}\right\rangle \le \left\langle N_{\Gamma} \right\rangle$. The singular corner contributions to $\left\langle N_{\Gamma} \right\rangle$ in Eq.(\ref{N_Gamma}), however, are
due to large clusters, which are present typically among the crossing clusters, too. Consequently
we expect the asymptotic form of the average entanglement entropy to be
\be
\left\langle {\cal S}_{\Gamma}\right\rangle =a L_{\Gamma}+b\ln L_{\Gamma} \;,
\label{S_Gamma}
\ee
where $a \le -f'_s(1)$ and $b=-C'_{\Gamma}(1)$. The leading term in Eq.(\ref{S_Gamma}) represents the so called
area law, to which there is a logarithmic correction, which is expected to be universal and given
in Eqs. (\ref{N_Gamma},\ref{cardy_peschel}) and (\ref{c'}).

\section{Numerical results}
\label{sec:numerical}

\begin{figure}[!ht]
\begin{center}
\includegraphics[width=3.2in,angle=0]{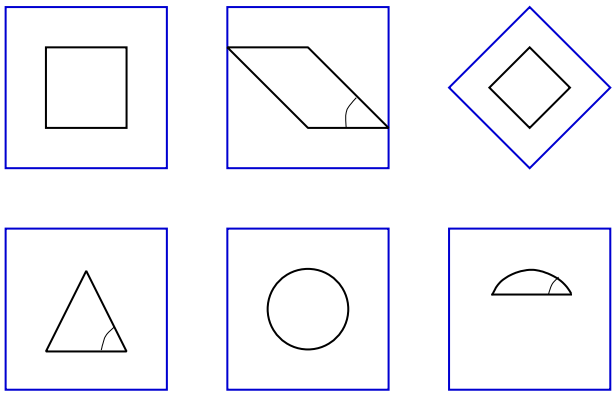}
\end{center}
\caption{
\label{fig_1} (Color online) Shapes of the subsystems used in the numerical calculation: square, sheared square, anisotropic square; equilateral triangle, circle and section of a circle. In a lattice we consider the nearest neighbor
bonds which are intersected by the contour of the subsystem and for each bond one of its sites
belongs to $\Gamma$ (depending on the local orientation of the surface).}
\end{figure}

We have performed large scale numerical calculations for site and bond percolations at the critical
point on the square lattice\cite{sedgewick}. The finite systems we have used have $L \times L$ sites with $L$ up to
8192 and with periodic boundary conditions.  The different shapes
of the subsystems used in the numerical calculations are illustrated in Fig.\ref{fig_1}.
In all cases the linear size of the subsystem is of ${\cal O}(L)$, so that $L_{\Gamma} \sim L$ and
in the following we shall use $L$ instead of $L_{\Gamma}$.
For a given shape of the subsystem and thus for its
boundary $\Gamma$ we have calculated the number of crossing clusters, ${\cal S}_{\Gamma}$, which is
then averaged:
i) for a given percolation sample over the positions of $\Gamma$ (typically $10^3$ positions), and
ii) over different samples. Typically we have used $10^5$ samples for each size, $L$, except the
largest ones, where we had at least $10^4$ samples. From the numerical data on $\left\langle {\cal S}_{\Gamma}\right\rangle$
we have deduced estimates for the prefactor $b$ by the so called \textit{difference approach}. Here we use the relation:
\be
\Delta {\cal S}_{\Gamma}(L)=2 \left\langle {\cal S}_{\Gamma}\right\rangle(L)-
\left\langle {\cal S}_{\Gamma}\right\rangle(2 L)=b \ln L + \text{const}\;.
\label{delta_S}
\ee
If $\Gamma$ has a special shape, such as a square or a sheared square, (see the three shapes in the
first row of Fig.\ref{fig_1}) we can calculate the
corner contribution to $\left\langle {\cal S}_{\Gamma}\right\rangle$ directly, by comparing
results obtained at different geometries. This type of \textit{geometrical approach}, which has been
used in Ref.~\onlinecite{kovacs_igloi12}, will be explained in more detail in the following subsection.

\subsection{Square subsystem}
\label{sec:square}

\begin{figure}[!ht]
\begin{center}
\includegraphics[width=3.2in,angle=0]{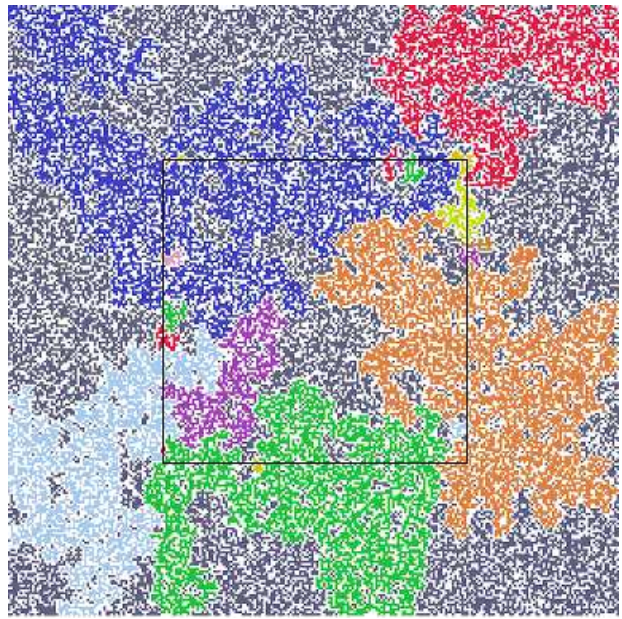}
\end{center}
\caption{
\label{fig_2} (Color online) Critical site percolation on a $L=128$ lattice. The subsystem is a square of size
$\ell=L/2=64$ and in this example $N_{\square}=26$ clusters intersect $\Gamma=\square$ and the number of crossing clusters is
$S_{\square}=22$.}
\end{figure}

The first geometry we consider for $\Gamma$ is a square of linear size $\ell=L/2$.
The cluster structure of critical site percolation is illustrated in Fig.\ref{fig_2}, in which one can
identify the clusters which intersect $\Gamma$ and also the crossing clusters. Here we use the
geometrical approach, in which the complete square is divided either to four neighboring squares
with $\ell=L/2$ or to orthogonal stripes of size $L/2 \times L$: altogether four stripes
due to the two different orientations. (This is illustrated in Ref.\onlinecite{kovacs_igloi12} in the
right panel of Fig.1.)
The two subsystems of different shapes have the same total boundary, however for stripes - due to periodic
boundary conditions - there is no corner contribution. This is then obtained from the difference:
\be
{\cal S}_{\square}^{cr}(L)={\cal S}_{\square}(L)-{\cal S}_{str}(L)\;.
\label{S_corner}
\ee
Note that by this geometrical method the corner contribution is calculated for each sample,
therefore the average values have considerably less noise, than by the difference method.

\begin{figure}[!ht]
\begin{center}
\includegraphics[width=3.2in,angle=0]{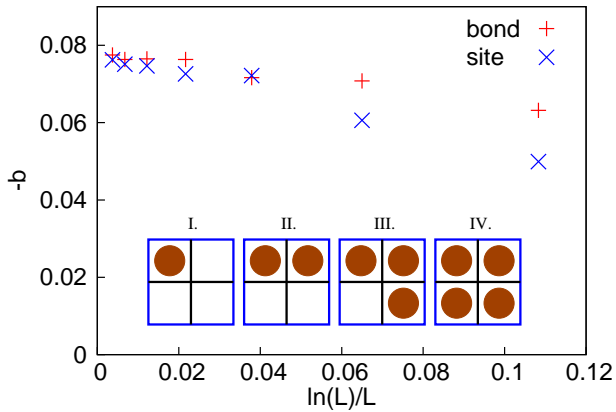}
\end{center}
\caption{
\label{fig_3} (Color online) Estimates for the prefactor of the corner contribution
$\left\langle {\cal S}_{\square}^{cr}\right\rangle$ of the square subsystem obtained from two-point fits. The asymptotic values as $L\to\infty$ are the same for site and bond percolation
and agree with the conformal prediction. Inset: Large scale topology of the clusters.
The square is divided into four quadrants and the different
topologies of clusters in the quadrants are illustrated: empty box $\to$ no site; filled box $\to$ at least one occupied
site.}
\end{figure}

Making use of the fact that asymptotically $\left\langle {\cal S}_{\square}^{cr}(L)\right\rangle
=b \ln L + \text{const}$, we have calculated 
effective, size-dependent estimates for the prefactor by
two-point fits, by comparing the average corner contributions in finite systems of size $L$ and $2L$. The
results are presented in Fig.\ref{fig_3} both for site and bond percolations. With increasing $L$ the
effective prefactors approach a common, universal limiting value of $b=-0.077(1)$. This is to be
compared with the conformal result in Eqs. (\ref{N_Gamma},\ref{cardy_peschel}) and (\ref{c'}) :
$-C'(1)=-5 \sqrt{3}/(36 \pi)=-0.07657$, thus the agreement is
satisfactory. We note that a previous numerical estimate in Ref.~\onlinecite{yu07} for smaller systems has obtained: $b=-0.06(1)$.
The corner contribution is related to the large scale topology of the clusters, as illustrated
in the inset of Fig.\ref{fig_3} and discussed in Sec.\ref{sec:corner_pr}.

\begin{figure}[!ht]
\begin{center}
\includegraphics[width=3.2in,angle=0]{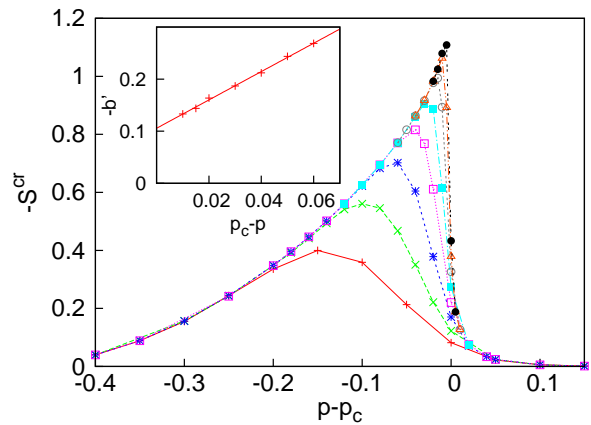}
\end{center}
\caption{
\label{fig_4} (Color online) $\Delta p=p-p_c$-dependence of the corner contribution to ${\cal S}_{\square}$
for different sizes, $L=16,32,\dots,2048$ from below for bond percolation. Close to $p_c$ the extrapolated curve has a
logarithmic singularity, see Eq.(\ref{S_corner_p}). In the inset the effective prefactors, $b'$, are shown as a function
of $|\Delta p|$. }
\end{figure}

We have also studied the $p$-dependence of the corner contribution to $\left\langle {\cal S}_{\square} \right\rangle$
outside the critical point by the geometrical approach. As can be seen in Fig.\ref{fig_4} ${\cal S}_{\square}^{cr}$
has a peak around $p=p_c$ and close to $p_c$ the extrapolated curve can be well described by the scaling result:
\be
{\cal S}_{\square}^{cr}(p) \simeq b' \ln(p_c-p)+const\;,
\label{S_corner_p}
\ee
where $b'=b \nu$, with $\nu=4/3$ being the correlation length critical exponent for percolation. Indeed, assuming
the form in Eq.(\ref{S_corner_p}) we have calculated effective, $p$-dependent prefactors by two-point fits,
which are shown in the inset of Fig.\ref{fig_4}. The extrapolated value for $p \to p_c$ ($0.106(10)$)
is consistent with the scaling prediction ($0.102$).

\subsection{Sheared square subsystem}
\label{sec:sheared_square}

\begin{figure}[!ht]
\begin{center}
\includegraphics[width=3.2in,angle=0]{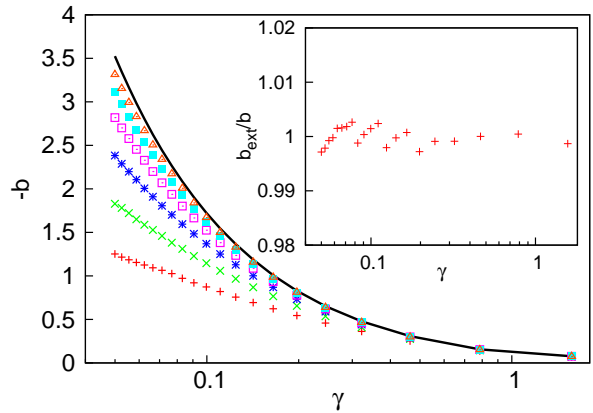}
\end{center}
\caption{
\label{fig_5} (Color online) Estimates for the prefactor $-b$ for sheared squares as a function of
$\ln(\gamma)$. The numerical results are for $L=128,~256,~\dots,~4096$ from below, the full line
represents the conformal result in Eq.(\ref{sheared}). In the inset the ratio of the extrapolated numerical data 
and the conformal result is given.}
\end{figure}

The square subsystem used in the previous subsection is sheared now to a parallelogram, having
the same surface: $\ell^2=L^2/4$ and its (smaller) angle is $\gamma$. The geometrical approach
to calculate the corner contribution to ${\cal S}_{\Diamond}^{cr}$ can be extended in this case
for specific values of $\gamma$ given by the condition: $\tan(\gamma)=1/n$, with $n=0,1,\dots,\infty$
being an integer. As shown in Fig.\ref{fig_5} with increasing $L$ up to $L=4096$ the numerical results approach
the conformal prediction:
\be
C'(1)=-\frac{c'(1)}{12}\left[4-\pi\left(\frac{1}{\gamma}+\frac{1}{\pi-\gamma}+\frac{1}{\pi+\gamma}+\frac{1}{2\pi-\gamma}\right)\right]\;.
\label{sheared}
\ee
Performing an extrapolation with an $\ln L/L$ correction term we have obtained an excellent agreement, as
shown in the inset of Fig.\ref{fig_5}. For other values of $\gamma$, which do not fit to the geometrical approach we
have made calculations by the difference method. Also in these cases the numerical results are found to
agree with the conformal prediction.

\subsection{Anisotropic percolation}
\label{sec:anisotrop}
Here we consider anisotropic bond percolation, in which the probabilities are $p_x$ (horizontally) and $p_y$ (vertically)
and the critical point is given by the condition: $p_x+p_y=1$. The system is a diagonally placed
square with $2L^2$ sites, and the subsystem is also a diagonally oriented square, having
$L^2/2$ sites and its boundary contains $4 \times L/2$ sites, see the third figure in the first row of Fig.\ref{fig_1}.
This anisotropic system with symmetric angles, $\pi/2$, is conjectured to be 
equivalent in the scaling limit to an isotropic system with asymmetric angles, $\gamma$ and $\pi-\gamma$, such that
\be
\dfrac{p_y}{p_x}=\dfrac{\sin((\pi-\gamma)/3)}{\sin(\gamma/3)}\;.
\label{p_y/p_x}
\ee
This follows from the requirement that there exists a discretely holomorphic observable\cite{ikhlef_cardy}. More recently Grimmett and Manolescu\cite{grimmett} have proved that many properties of the scaling limit are the same if a more general inhomogenously anisotropic lattice is embedded in the plane according to this prescription.

\begin{figure}[!ht]
\begin{center}
\includegraphics[width=3.2in,angle=0]{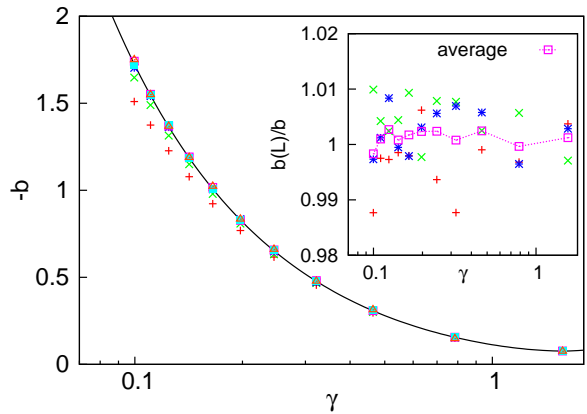}
\end{center}
\caption{
\label{fig_6} (Color online) Estimates for the prefactor $-b$ for anisotropic percolation
for finite systems with $L=32,~64,~\dots,~1024$ from below. The opening angle of the
equivalent sheared square, $\gamma$ is given by Eq.(\ref{p_y/p_x}) and the full line
represents the conformal result. In the inset the ratio of the finite-size results and
the conformal results are given.}
\end{figure}

In the anisotropic system we have chosen the probabilities in such a way, that the corresponding angle, $\gamma$,
has satisfied the condition: $\tan(\gamma)=1/n$, with $n=0,1,\dots,\infty$. In this way we could directly
compare the results for anisotropic systems with those obtained for sheared squares in Sec. \ref{sec:sheared_square}.
The numerical results for the prefactor obtained on finite systems up to $L=1024$ are are shown in Fig.\ref{fig_6}.
For the three largest systems there is no systematic size-dependence of our data, the error is purely statistical
and the numerical results agree well with the conformal prediction.

\subsection{Corner probability}
\label{sec:corner_pr}
For the (sheared) square subsystem we used the geometrical method, in which
the $L \times L$ system is divided for four $L/2 \times L/2$ squares and for four $L/2 \times L$
stripes and the difference in the number of crossing clusters in Eq.(\ref{S_corner}) is just the corner contribution.
Having a general (connected) cluster it could occupy different parts of the four quadrants and its topology
could be of four different types as illustrated in the inset of Fig.\ref{fig_3}. Among these the I, II and IV type
of topology gives identical contribution both for stripes and squares. Clusters
with topology III, however, are crossing clusters for all the four possible stripes, but these are crossing clusters only
for three out of four squares. (In the left-bottom square there is no crossing.) Let us denote the occurrence
probability of type III cluster as $P_3(L)$, which is given by the ratio of such clusters in a $L \times L$ square,
which have points in three quadrants, but have no point in the fourth one. As argued above $P_3(L)$ is proportional
to the corner contribution of crossing squares, more precisely:
\be
P_3(L)=\dfrac{1}{4} {\cal S}_{\square}^{cr}(L)=-\dfrac{1}{4}C'(1)\ln L +\text{const.}\;.
\label{P_3}
\ee
This results is valid for sheared squares with angle $\gamma$, as well as for anisotropic percolation
with a square subsystem. In these cases the appropriate results of ${\cal S}_{\square}^{cr}(L)$ have to
be used.
Interestingly this corner probability has a logarithmic $L$-dependence and its prefactor is known exactly.

\subsection{Other subsystem geometries}
\label{sec:Other}
We have also studied subsystems with different geometries: equilateral triangle, circle and section of a circle,
these are illustrated in the second row of Fig.\ref{fig_1}.
In these cases the largest linear scale of the subsystem is fixed to $L/2$, while an angle was varied. In the
calculations the difference approach in Eq.(\ref{delta_S}) has been used.

\subsubsection{Equilateral triangle}
\begin{figure}[!ht]
\begin{center}
\includegraphics[width=3.2in,angle=0]{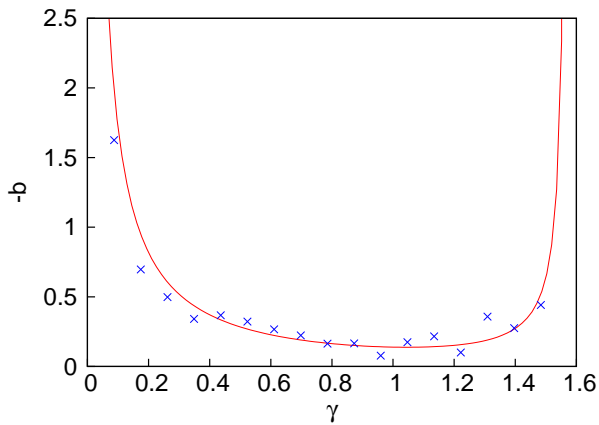}
\end{center}
\caption{
\label{fig_7} (Color online) Prefactor of the logarithm for an equilateral triangle as a function
of the base angle $\gamma$ compared with the conformal result in Eq.(\ref{triangle}).
The numerical results are extrapolated up to $L=4096$ for site percolation.
The statistical error is larger, than the difference between the conformal and the numerical results.}
\end{figure}

For an equilateral triangle with a base angle $\gamma$ the extrapolated results for $-b$
are shown in Fig.\ref{fig_7} , which is compared with the conformal prediction:
\be
C'(1)=-\frac{c'(1)}{24}\left[6-\pi\left(\frac{2}
{\gamma}+\frac{1}{\pi-2\gamma}+\frac{1}{\pi+2\gamma}+\frac{2}{2\pi-\gamma}\right)\right]\;.
\label{triangle}
\ee
As seen in Fig. \ref{fig_7} there is a satisfactory agreement, although the statistical error of the numerical
results is comparatively large, in particular for small and large angles.

\subsubsection{Circle and section of a circle}

\begin{figure}[!ht]
\begin{center}
\includegraphics[width=3.2in,angle=0]{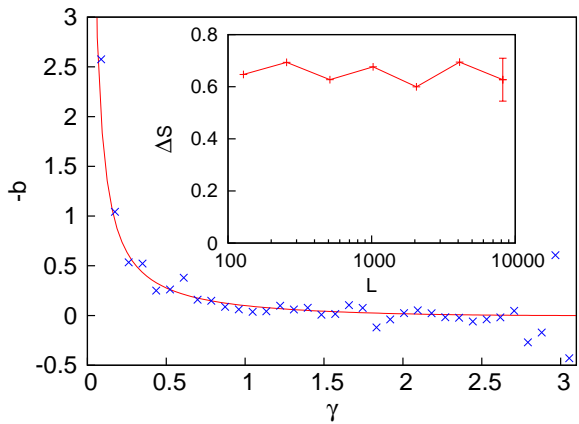}
\end{center}
\caption{
\label{fig_8} (Color online) Prefactor of the logarithm $-b$ for a section of a circle as a function of the
angle $\gamma$. The full line represents the conformal result in Eq.(\ref{section}). The differences
between the conformal and numerical results are of the order of the statistical error. In the inset the
corner contribution is shown for a circle: there is no systematic radius dependence.
The numerical results are extrapolated up to $L=4096$ (in the main figure) and
up to $L=8192$ (in the inset) for site percolation.}
\end{figure}

According to the conformal results, there is no logarithmic correction to $S_{\bigcirc}$ for a circle shaped subsystem.
Indeed the numerical results  in the inset of Fig.\ref{fig_8} are in agreement with this statement: $\Delta S_{\bigcirc}$
approaches a finite limiting value of $0.65(5)$.

In contrast, for a section of a circle with an angle $0 < \gamma < \pi$ there is a logarithmic correction, the
prefactor of which is given by conformal invariance:
\be
C'(1)=-\frac{c'(1)}{12}\left[2-\pi\left(\frac{1}{\gamma}+\frac{1}{2\pi-\gamma}\right)\right]\;
\label{section}
\ee
The numerical results in Fig.\ref{fig_8} are in agreement with the conformal prediction, although the statistical
error of the numerical results is comparatively large.

\subsection{Line segment}
\label{sec:line}
According to the derivation in Sec.\ref{sec:Potts} $\Gamma$ does not have to be a closed curve. For example
it could be a straight line of length $\ell$. In that case we have only a corner contribution from two
exterior angles, each $\gamma=2 \pi$, so that $C_{\Gamma}(Q)=-c(Q)/8$.

To study this problem we can use the geometrical approach, when the line is oriented parallel with one of the axes of the
square lattice. Then the corner contribution is related to the difference
between the cluster numbers obtained for a
periodic line of length $L$ and that of two segments of lengths $\ell=L/2$. In
this case the corner contribution is simply half of the number of common clusters
between the two line segments.

\begin{figure}[!ht]
\begin{center}
\includegraphics[width=3.2in,angle=0]{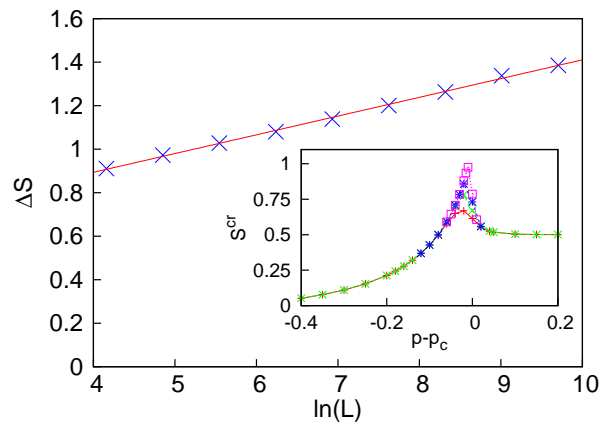}
\end{center}
\caption{
\label{fig_9} (Color online) $\Delta S$ for the line segment obtained by the difference approach for different
lengths, $L$. In the inset the $p$-dependence of $S^{cr}$ is shown close to the critical point
for finite systems with $L=64, 128, 256$ and $512$ from below.}
\end{figure}

In the numerical calculation we have used the difference approach for a line segment of length $\ell=L/2$, which is
placed in random positions and random orientations with respect to the axes of the periodic square lattice of size $L \times L$.
The average of $\Delta S_{\arrowvert}(L)$ is shown in Fig. \ref{fig_9} as a function of $\ln L$. The logarithmic dependence is clearly visible with a prefactor estimated as $b=0.086(1)$, which agrees fairly well with the conformal prediction: $-C'_{\Gamma}(1)=\frac{5\sqrt{3}}{32\pi}\approx 0.08615$. We have also checked the $p$-dependence of
$S^{cr}_{\arrowvert}(p)$, which is shown in the inset of Fig. \ref{fig_9}.
A singularity is developed at the critical point as $L \to \infty$.

We note that if the line segment is put to the open boundary of the system, then the prefactor
is different, being $b=\frac{\sqrt{3}}{4\pi}\approx 0.1378$ as obtained by conformal invariance\cite{cardy01,yu07} and
numerically $b=0.15(2)$ \cite{yu07}.

\section{Discussion}
\label{sec:disc}
In this paper we  considered the logarithmic terms in the mean number $N_\Gamma$ of clusters in critical percolation which intersect a curve $\Gamma$ in the cases when it has sharp corners or end points. We have shown that the Cardy-Peschel result\cite{cardypeschel} can be simply applied to compute the universal coefficients of these logarithmic terms, and that accurate numerical estimates agree very well with this, for a variety of shapes for $\Gamma$. We also considered anisotropic bond percolation on the square lattice and showed that if Eq.~(\ref{p_y/p_x}) is used to deform the lattice then the predictions again agree with numerics if the correct effective corner angle is used. We have also pointed out a relation between the corner
contribution to $N_{\Gamma}$ and the statistics of cluster shapes.

Our study is related to recent investigations of shape dependent terms of different thermodynamic
quantities, mainly the free energy, of $2d$ critical systems. For critical percolation, however, there is no
logarithmic corner contribution to the free energy, since the central charge is $c(Q=1)=0$. In the latter problem
the corner contribution to the cluster numbers and its higher moments are of interest, which are related
to derivatives of $c(Q)$ at $Q=1$.

The excellent agreement of the numerical data with the theory serves as confirmation that critical percolation is indeed conformally invariant, of the Coulomb gas predictions for $c'(Q=1)$, and of the formula (\ref{p_y/p_x}). Extensions are obviously possible: for example by taking further derivatives with respect to $Q$ we can find results for the corner contributions to the higher moments of the distribution of the cluster numbers $N_\Gamma$. We note that recently\cite{Vasseur} an explicit example has been found of a correlation function in percolation which contains a multiplicative logarithm. As with our result, this may be understood\cite{cardylogs} from the necessity of having to take a suitable derivative with respect to $Q$ at $Q=1$ in the Potts model.

\begin{acknowledgments}
This work has been supported by the Hungarian National Research Fund under Grants
No. OTKA K75324 and K77629. This work has been partly
done when two of the authors (J. C. and F. I.) were guests of the Galileo Galilei Institute in Florence
whose hospitality is kindly acknowledged.
\end{acknowledgments}

\end{document}